\newcommand{\subtil}[1]{\makebox[0mm][l]{\raisebox{-3mm}[3mm][-2mm]{$\
\tilde{}$}}#1}
\newcommand{\One}{{\bf 1}}
\newcommand{\ag}{\alpha}
\newcommand{\bg}{\beta}
\newcommand{\cg}{\gamma}
\newcommand{\dg}{\delta}
\newcommand{\sg}{\sigma}
\newcommand{\m}{\mu}
\newcommand{\n}{\nu}
\newcommand{\eg}{\epsilon}

\newcommand{\Sg}{\Sigma}

\newcommand{\tilE}{\tilde{E}}

\newcommand{\di}{\partial}
\newcommand{\be}{\begin{equation}}
\newcommand{\ee}{\end{equation}}
\newcommand{\bearr}{\begin{eqnarray}}
\newcommand{\eearr}{\end{eqnarray}}
\newcommand{\implies}{\Rightarrow}
\newcommand{\Real}{{\bf R}}

\newcommand{\QED}{\rule{1mm}{3mm}}

\documentstyle[12pt,psfig]{article}
\begin{document}

\title{Classical Euclidean general relativity from ``left-handed area 
= right-handed area"}
\author{Michael P. Reisenberger\thanks{e-mail: miguel@fisica.edu.uy}\\
 Instituto de F\'{i}sica, Universidad de la Rep\'{u}blica\\
      Trist\'{a}n Narvaja 1674, 11200 Montevideo, Uruguay}
\date{April 14, 1998}
\maketitle

\begin{abstract}
	A classical continuum theory corresponding to Barrett and Crane's
model of Euclidean quantum gravity is presented. The fields in this classical
theory are those of $SO(4)$ BF theory, a simple topological theory of an
$so(4)$ valued 2-form field, $B^{IJ}_{\m\n}$, and an $so(4)$ connection.
The left handed (self-dual) and right handed (anti-self-dual) components of
$B$ define a left handed and a, generally distinct, right handed area for
each spacetime 2-surface. The theory being presented is obtained by adding 
to the BF action a Lagrange multiplier term that enforces the constraint that
the left handed and the right handed areas be equal. It is shown that 
Euclidean general relativity (GR) forms a sector of the resulting theory.
The remaining three sectors of the theory are also characterized and it is 
shown that, except in special cases, GR canonical initial data is sufficient 
to specify the GR sector as well as a specific solution within this sector.

Finally, the path integral quantization of the theory is discussed at a formal
level and a hueristic argument is given suggesting that in the semiclassical 
limit the path integral is dominated by solutions in one of the non-GR sectors,
which would mean that the theory quantized in this way is not a quantization of 
GR.
\end{abstract}

\section{Introduction}

The Palatini action for general relativity (GR) may be written in the 
form\footnote{
The wedge product of two forms, $a_{\ag_1 ...\ag_m}$ and $b_{\bg_1 ...\bg_n}$,
is defined as
\be
[a \wedge b]_{\ag_1...\ag_m\bg_1...\bg_n} = a_{[\ag_1...\ag_m}
b_{\bg_1...\bg_n]}.
\ee
The integral over a $d$ dimensional submanifold, $s$, of a $d$-form, $f$, 
living on a manifold $M$ is defined as
\be
\int_s f = \int f_{\ag_1 ... \ag_d} \frac{\di x^{[\ag_1}}{\di\sg^1} ...
\frac{\di x^{\ag_d]}}{\di\sg^d}	d^d\sg,
\ee
where $\sg^1,...,\sg^d$ are coordiantes on $s$, and $x^{\ag}$ are coordinates
on $M$.} 
\be		\label{Palatini}
I_{GR} = \int \eg_{IJKL}\: e^I \wedge e^J \wedge R^{KL},
\ee
where $e^I_\m$ is the vierbein, $R^{KL}_{\m\n}$ is the curvature of the spin connection, and 
the capital latin indices $I, J, ... \in \{ 0,1,2,3\}$ are vector indices under the 
frame rotation group - $SO(4)$ in the case of Euclidean GR which we shall consider.

(\ref{Palatini}) is just the $SO(4)$ BF\footnote{%
The BF theory for (internal) gauge group $G$ is defined by the action $\int tr[B\wedge F]$ 
when $F$ is the curvature and $B$ is a 2-form field taking values in the adjoint represtation
of the algebra of $G$, just like $F$. The trace is taken in this adjoint representation. See
\cite{SchBF}, \cite{HorBF}.}
 action, $\int B_{KL}\wedge R^{KL}$ with the restriction
that 
\be		\label{vierB}
B_{KL} = \eg_{IJKL}\ e^I\wedge e^J.
\ee
Barrett and Crane \cite{BC} have proposed a model of simplicial Euclidean quantum gravity in
which the amplitude for any history is its amplitude in the simplicial BF theory of Ooguri
\cite{Ooguri},\footnote{
Ooguri's model was of $SU(2)$ BF theory. Here the (trivial) extension
to $SO(4)$ BF theory is used. Barrett and Crane also consider regulating their model by 
q-deforming the $so(4)$ gauge algebra, which presumably corresponds to introducing a cosmological
constant. In this case the Ooguri model is replaced by the Crane-Yetter model \cite{BC}.}
but the histories are restricted by constraints designed to enforce a simplicial version 
of (\ref{vierB}). In fact these constraints have several branches of solutions besides 
those corresponding to (\ref{vierB}) \cite{Baez}. 

It turns out \cite{Urbantke,CDJM,Bengtsson,Rei95} that when (\ref{vierB}) holds the geometrical 
(gauge invariant) information 
contained in the vierbein $e$, is equivalent to the gauge invariant information in the self-dual, 
or ``left handed", part, $B^+$, of $B$, as well as that in the anti-self-dual, or ``right-handed", 
part $B^-$. Since $B^+$ and $B^-$ are well defined for any $B$ they can be used to define a left 
handed and a right handed spacetime geometry whether (\ref{vierB}) holds or not.
Baez \cite{Baez} has pointed out that
the Barrett-Crane constraints imply that the left and right geometries of 3-surfaces
embedded in spacetime are equal. In particular this implies that the areas of spacetime 
2-surfaces computed from $B^+$ and from $B^-$ are equal. As the author hopes to elaborate
in a future publication, the equality of self-dual and
anti-self-dual areas is in fact the essence of the Barrett-Crane constraints.

Here we will examine the analogous {\em classical continuum} theory, defined by the stationary 
points of the BF action with respect to variations that preserve the equality of self-dual and 
anti-self-dual areas. We will find that a branch of these stationary points reproduce GR.
An action for this theory is constructed in \S \ref{setup}, and its several 
branches of solutions are classified and studied in section \S \ref{solns}.
Finally, \S \ref{comment} is a hueristic comment on the quantization of the theory, and \S \ref{conclusion} summarizes the conclusions.

\section{Setting up the theory}	\label{setup}

Let's begin by briefly reviewing self-duality and the decomposition of $SO(4)$ into left and 
right handed factors. The dual of an antisymmetric tensor $a^{IJ}$ is defined as\footnote{
$SO(4)$ tensor indices are raised with the $SO(4)$ metric $\dg^{IJ}$, i.e. the Kronecker delta.}
\be
a^{\star\,IJ} = \frac{1}{2}\eg^{IJ}{}_{KL}a^{KL}.
\ee
Any antisymmetric tensor $a^{IJ}$ can be split into a self-dual component $a^+$ 
satisfying $a^{+} = a^{+\star}$, and an anti-self-dual component $a^-$ 
satisfying $a^{-}= - a^{-\star}$, according to
\begin{equation}	\label{SDdef}
a^{\pm} = \frac{1}{2}[ a \pm a^\star ].
\end{equation}

$SO(4)$ factors into a ``left-handed" $SU(2)$ and a ``right-handed" $SU(2)$:
\be		\label{so4decomp}
SO(4) = SU(2)_L \otimes SU(2)_R.
\ee
Self-dual tensors transform non-trivially only under $SU(2)_L$. Under $SU(2)_R$ they are constant.
For this reason they are also called ``left-handed" tensors. Similarly anti-self-dual
tensors transform non-trivially only under $SU(2)_R$, and are called ``right-handed" tensors.
(Anti-)self-dual tensors have only three independent components, which will be taken to be 
\begin{equation} \label{SDcomponents}
a^{\pm i} = \pm 2 a^{\pm 0i}.
\end{equation}
(Lower case latin indices run over $\{1,2,3\}$).
These components transform as adjoint vector ($=SO(3)$ vector) components, under $SU(2)_L$ in 
the case of a left-handed tensor, and under $SU(2)_R$ in the case of right-handed tensors.

It is easy to verify from (\ref{SDdef}) and (\ref{SDcomponents}) that for any two antisymmetric tensors
$a^{IJ}$ and $b_{KL}$
\be
a^{IJ} b_{IJ} = a^{+\,i}b^+_i + a^{-\,i}b^-_i.
\ee
It follows that the $SO(4)$ BF action, $\int B_{IJ}\wedge R^{IJ}$ can be written as the 
difference of a left handed and a right handed piece:\
\be		\label{BF2}
I_{BF} = 2 \int \Sg_i^+ \wedge R^{+\,i}  -  \Sg_i^- \wedge R^{-\,i}.
\ee
with $\Sg\pm_i = 2B^{\pm\,0i}$. This reduces to the Palatini action for GR when a condition
equivalent to (\ref{vierB}) is imposed:
\be		\label{metric_form}
\Sg^{\pm}_i = \pm 2 E^{\pm\,0i} = \pm e^0 \wedge e^i 
+ \frac{1}{2} \eg_{ijk} e^j \wedge e^k,
\ee
where $E^{IJ} = e^I \wedge e^J$. It follows from the factorization (\ref{so4decomp}), 
or by simple algebra, that the left and right handed components, $R^+$ and $R^-$,
of the $SO(4)$ curvature $R$ are the curvatures of, $A^+$ and $A^-$, the $SU(2)_L$ 
and $SU(2)_R$ components, respectively, of the spin connection. Since any pair
of $SU(2)$ connections define an $SO(4)$ connection via (\ref{so4decomp}), there are
no restrictions $A^\pm$. $I_{BF}$ is thus the difference of two independent $SU(2)$
BF actions. GR is obtained by coupling the left and right handed fields via the
constraint (\ref{metric_form}).

Perhaps surprisingly, when (\ref{metric_form}) holds both $\Sg^+$ and $\Sg^-$ contain enough 
information to compute the
geometry of spacetime. The metric can be determined from either $\Sg^+$ or $\Sg^-$ 
\cite{Urbantke,CDJM,Bengtsson,Rei95}. Our principal interest will be areas. 

Just as the length of a curve in spacetime can be written in terms of the usual metric, 
the area of a 2-surface can be written in terms of an ``area metric":
\be
\mbox{Area}(\sg) = \int_\sg \sqrt{m_{\ag\bg\cg\dg} t^{\ag\bg} t^{\cg\dg}} du^1 du^2,
\ee
where $\sg$ is any smooth 2-surface, $(u^1, u^2)$ are coordinates on $\sg$, 
$t^{\ag\bg} = 2 \frac{\di x^{[\ag}}{\di u^1} \frac{\di x^{\bg]}}{\di u^2}$ is the tangent bivector
of $\sg$ (the $x^\cg$ are spacetime coordinates), and 
\be
m = \frac{1}{2} E^{IJ} \otimes E_{IJ}
\ee
is the area metric.

When (\ref{metric_form}) holds 
this area metric can be written in terms of either $\Sg^+$ or $\Sg^-$. It is equal to both the
left area metric, $m^+$, and the right area metric, $m^-$, defined by
\be
m^{\pm} = \Sg^{\pm\,i}\otimes \Sg^\pm_i - \frac{1}{4} V^\pm,
\ee
with
\be
V^\pm = \pm 4 \Sg^{\pm\,i}\wedge \Sg^\pm_i.
\ee
$V^+$ and $V^-$ both equal the 4-volume form $V = e^0\wedge e^1\wedge e^2\wedge e^3$.

If $\Sg^+$ and $\Sg^-$ are not required to satisfy (\ref{metric_form}) the left and right
areas and 4-volumes are still defined, though they need not agree. $m^+$ and $V^+$ define 
a ``left" geometry, while $m^-$ and $V^-$ defines a, generally distinct,
``right" geometry.
\footnote{One can also define left and right metrics for lengths from $\Sg^+$ and $\Sg^-$
via Urbantke's formula \cite{Urbantke,Bengtsson}
\be
[\sqrt{g}g_{\m\n}]^\pm = \pm[8\eg^{\ag\bg\cg\dg}\Sg^1_{\m\ag}\Sg^2_{\bg\cg}\Sg^3_{\dg\n}]^\pm.
\ee
However the areas defined by 
these length metrics will generally be different from those defined by the left and right 
area metrics used here.}

Suppose the left and right areas are {\em required} to be equal. What does this tell us about
the form of $\Sg^\pm$? 
It turns out that when the average 4-volume form $\frac{1}{2}(V^+ + V^-) \neq 0$ the equation
$m^+ = m^-$ has four isolated branches of solutions\footnote{
The corresponding four branches of solutions to Barrett and Crane's essentially equivalent 
equation on left and right handed area bivectors of the the 2-simplices of a 4-simplex, are
already given in \cite{Baez}.}
one of which is the vierbein form (\ref{metric_form}). In addition there are two branches of
solutions with $\frac{1}{2}(V^+ + V^-) = 0$.

Thus standard GR can be obtained as a sector of the solutions of the theory defined by
\bearr 
I_\psi & = & \int \Sg^+_i\wedge F^{+i} - \Sg^-_i \wedge F^{-i} \\
   && \ \ + \int \psi^{\ag\bg\cg\dg} [\Sg^+_{i\,\ag\bg}\Sg^{+\,i}_{\cg\dg} - 
												\Sg^-_{i\,\ag\bg}\Sg^{-\,i}_{\cg\dg}] d^4x, 
\label{LR_action}
\eearr
where $\Sg^+$, and $\Sg^-$ are independent triplets of 2-form fields, 
$F^+$ and $F^-$ are the curvatures of two independent $SU(2)$ connections, and $\psi$ is a Lagrange
multiplier of density weight one, which satisfies $\psi^{\ag\bg\cg\dg} = 
\psi^{\cg\dg\ag\bg} = - \psi^{\bg\ag\cg\dg}$ and 
$\psi^{[\ag\bg\cg\dg]} = 0$. The last condition ensures that stationarity with respect to $\psi$
requires only the equality of the left and right area metrics, $m^\pm_{\ag\bg\cg\dg}$, which 
have no totally antisymmetric part, and leaves the totally antisymmetric parts of 
$\Sg^\pm\otimes\Sg^\pm$, which form $V^\pm$, unconstrained.

The field equations implied by the stationarity of $I_\psi$ are
\bearr
0 & = & D^\pm \wedge \Sg^\pm 	\label{curl}	\\
0 & = & F^{\pm\,i}_{\m\n} + \frac{1}{4} \eg_{\m\n\ag\bg}\psi^{\ag\bg\cg\dg}\Sg^{\pm\,i}_{\cg\dg}
\label{Einst}	\\
0 & = & m^+ - m^-.	\label{lrareas}
\eearr

$m^+ - m^-$ has 20 linearly independent components,
so $m^+ - m^- = 0$ seems to provide just the right number of constraints to reduce the 36 degrees of
freedom of $(\Sg^+,\Sg^-)$ to the 16 degrees of freedom of the vierbein $e^I_\m$.
A more careful study yields
\newline
\newline
\noindent {\bf Theorem}:

1) If $m^+ = m^-$ and $V^+ + V^- \neq 0$ then there exists a non-degenerate, real, vierbein
$e^I_\m$ and signs $s_1, s_2 \in \{1, -1\}$ such that
\bearr
\Sg^+_i & = & s^+ [ e^0 \wedge e^i + \frac{1}{2} \eg_{ijk} e^j\wedge e^k ] \label{tetform+} \\
\Sg^-_i & = & s^- [- e^0 \wedge e^i + \frac{1}{2} \eg_{ijk} e^j\wedge e^k ] \label{tetform-} \\
V^+ = \:V^- & = & e^0\wedge e^1 \wedge e^2 \wedge e^3. \label{tetformV}
\eearr

2) If $m^+ = m^-$ and $V^+ + V^- = 0$ then there exists $T \in SO(3)$ and $s \in \{1,-1\}$
such that
\be		\label{degform}
\Sg^-_j = s T_j{}^i \Sg_i^+.
\ee

\noindent{\em Proof}: When $m^+ - m^- = 0$
\be
\Sg^+_i \otimes \Sg^{+\,i} - \Sg^-_i \otimes \Sg^{-\,i}  = 
4 (V^+ + V^-)
\ee
because only the totally antisymmetric part of the expression on the left is non-zero.

Now let's consider 2): $V^+ + V^- = 0$ so
\be		\label{pm_eq}
\Sg^+_i \otimes \Sg^{+\,i} = \Sg^-_i \otimes \Sg^{-\,i}.
\ee
The $\Sg^{\pm\,i}$ define linear functions $v^{\pm\,i} = \Sg^{\pm\,i}_{\ag\bg} v^{\ag\bg}$ 
on the space $\cal A$ of rank two antisymmetric spacetime tensors $v^{\ag\bg}$. In terms of these
linear functions (\ref{pm_eq}) becomes $v^+_i v^{+\,i} = v^-_i v^{-\,i}$.
Now, $w^{+\,i} = u^{+\,i}$ implies $0 = [w-u]^+_i [w-u]^{+\,i} = [w-u]^-_i [w-u]^{-\,i}$. This,
in turn, implies that $w^{-\,i} = u^{-\,i}$. Obviously the converse also holds, so $v^{+\,i}$
is an invertible function of the $v^{-\,i}$. It is easy to see from the linearity of $v^{\pm\,i}$
that $v^{+\,i}$ and $v^{-\,i}$ are related by a linear transformation. Finally, 
$v^+_i v^{+\,i} = v^-_i v^{-\,i}$ implies that this transformation is orthogonal:
\be
\exists O \in O(3)\ \mbox{such that}\ v^{+\,i} = O^i{}_j v^{-\,j}.
\ee
This equation holds for all $v \in \cal A$, so it actually holds for the $\Sg^{\pm\,i}$.
If the orthogonal transformation is written as a product $O = sT$ of $T \in SO(3)$ and
a sign factor $s\in \{1,-1\}$ the result is
\be
\Sg^-_j = s T_j{}^i \Sg_i^+.
\ee 

Now let's prove part 1), which concerns the case $V^+ + V^- \neq 0$. The totally antisymmetric 
tensor
\be		\label{nondeg_metric}
\Sg^+_i \otimes \Sg^{+\,i} - \Sg^-_i \otimes \Sg^{-\,i} = 4 (V^+ + V^-).
\ee
defines a metric on $\cal A$ which is proportional to $\eg_{\ag\bg\cg\dg}$. This metric,
like $\eg$ is non-degenerate with signature $1,1,1,-1,-1,-1$. In order that the left
side of (\ref{nondeg_metric}) reproduce this non-degenerate metric it is necessary that the
six $\Sg^\pm_i$ be linearly independent. Consequently there exists a basis 
$\Lambda^{\pm\,\ag\bg}_i$ of $\cal A$ such that
\bearr
\Sg^+_{i\,\ag\bg}\Lambda^{+\,\ag\bg}_j = \dg_{ij} && \Sg^-_{i\,\ag\bg}\Lambda^{+\,\ag\bg}_j = 0 \label{cobasis}\\
\Sg^+_{i\,\ag\bg}\Lambda^{-\,\ag\bg}_j = 0 && \Sg^-_{i\,\ag\bg}\Lambda^{-\,\ag\bg}_j = \dg_{ij}.
\eearr
Contracting (\ref{nondeg_metric}) with $\Lambda^+_j$ one sees that
\be
\Sg_{j\,\ag\bg}^+ = \frac{1}{3!}(V^+ + V^-)_{\m\n\sg\tau}\eg^{\m\n\sg\tau}\eg_{\ag\bg\cg\dg} 
\Lambda^{+\,\cg\dg}_j.
\ee
When this is substituted back into (\ref{cobasis}) one finds that
\be		
\Sg^+_i \wedge \Sg^+_j = \dg_{ij}\frac{2}{3}(V^+ + V^-).
\ee
It follows that 
\be				\label{SG+wedgeSg+}
\Sg^+_i \wedge \Sg^+_j = \dg_{ij}\frac{1}{3}\Sg^+_k \wedge \Sg^{+\,k}
\ee
and, from the definition $V^+ = \frac{1}{4}\Sg^+_k \wedge \Sg^{+\,k}$, that 
$V^+ = \frac{1}{2}(V^+ + V^-) = V^-$.

By a well known result 
(\ref{SG+wedgeSg+}) and the fact that $\Sg^+_k \wedge \Sg^{+\,k} = 2(V^+ + V^-) \neq 0$ imply
that there exists a real vierbein $\bar{e}^I_\m$ and $s^+ \in \{ 1, -1\}$ such that
\be
\Sg^+_i = s^+ [\bar{e}^0 \wedge \bar{e}^i + \frac{1}{2} \eg_{ijk} \bar{e}^j\wedge \bar{e}^k]
\ee
and $\bar{e}^0\wedge \bar{e}^1\wedge\bar{e}^2\wedge\bar{e}^3 = V^+ \neq 0$
\cite{Israel}\cite{Plebanski77}\cite{CDJM} (see \cite{Rei95} appendix B for 
a proof in more or less the present notation).

Using this vierbein we may define a basis of anti self-dual 2-forms: 
${\Sg'}^-_i = -\bar{e}^0 \wedge \bar{e}^i + \frac{1}{2} \eg_{ijk} \bar{e}^j\wedge \bar{e}^k$. Then
${m'}^- = m^+ = m^-$ and ${V'}^- = V^+ = V^-$, so
\be
\Sg^-_i \otimes \Sg^{-\,i} - {\Sg'}^-_i \otimes {\Sg'}^{-\,i} = 0.
\ee
By 2), which is already proven, this implies that
\be
\Sg^-_i = s^- T_i{}^j {\Sg'}^-_j
\ee
for some $T\in SO(3)$ and $s^- \in \{1,-1\}$. Now let $A = T\otimes \One$ be the $SO(4)$ 
transformation composed of left handed factor $T$ and right handed factor $\One$. Then
$e^I_\m = A^I{}_J \bar{e}^J_\m$ and $(s^+, s^-)$ satisfy the equations (\ref{tetform+}), 
(\ref{tetform-}), and (\ref{tetformV}).\QED
\newline
\newline
The solutions of $m^+ = m^-$ can be characterized as follows: Given {\em any} triplet
of 2-forms $\Sg^+_i$, $m^+ = m^-$ can be satisfied by choosing $\Sg^-_i = O_i{}^j\Sg^+_j$
where $O \in O(3)$. For those special $\Sg^+_i$ that satisfy $\Sg^+_i\wedge\Sg^+_j \propto \dg_{ij}$,
with $\Sg^+_k \wedge \Sg^{+\,k} \neq 0$, there are four additional branches of solutions
(described in part 1) of the Theorem). In one of these branches of special solutions
$\Sg^+$ and $\Sg^-$ are of the vierbein form (\ref{metric_form}).

Now suppose we look for the stationary points of the action $I_{\psi}$ by first restriciting
attention to field configurations that extremize the action with respect to $\psi$ and 
then looking for stationary points among such fields. The solutions we are looking for 
are then stationary points of a difference of two BF actions, 
$\int \Sg^+_i\wedge F^{+\,i} - \Sg^+_i\wedge F^{+\,i}$, under variations subject to the 
constraint $m^+ = m^-$. For each branch of solutions to $m^+ = m^-$ an action $I_{branch}$ 
may be obtained by substituting the general solution in that branch into $I_{\psi}$.
The stationary points of $I_{\psi}$ are stationary points of one or another of the 
branch actions. Conversely, 
if the fields $(\Sg^+,\Sg^-)$ of a stationary point of a branch action belong only to that branch, 
i.e do not lie in the intersection of two branches of solutions to $m^+ = m^-$, then this
stationary point is also a stationary point of $I_\psi$. Stationary points of $I_\psi$ in the 
intersection of several branches must be stationary points of the actions of all these branches.

$I_{\psi}$ gives rise to four inequivalent branch actions
\bearr
I_{GR} & = & \int \eg_{IJKL}\ e^I \wedge e^J \wedge R^{KL} \\
I_1 & = & 2 \int e_I \wedge e_J \wedge R^{IJ}			\\
I_2 & = & \int \Sg_i \wedge [ F^{+\,i} + F^{-\,i} ] = \int \eg_{IJKL}\ a^I B^J \wedge R^{KL} \\
I_3 & = & \int \Sg_i \wedge [ F^{+\,i} - F^{-\,i} ] = 2 \int a_I B_J \wedge R^{IJ},
\eearr
where $R$ is an $SO(4)$ curvature, $\Sg_i$ is a triplet of 2-forms, $a$ is an $SO(4)$ vector,
and $B$ is an $SO(4)$ vector and a 2-form.

$I_{GR}$ corresponds to $\Sg^\pm$ of the vierbein form, which constitute the branch of solutions 
(\ref{tetform+}), (\ref{tetform-}) to $m^+ = m^-$ with $s^+ = s^- = 1$. $R$ appearing in $I_{GR}$,
and all the other branch actions, is the $SO(4)$ curvature defined by $F^+$ and $F^-$ appearing
in the general form, $I_\psi$, of the action. That is, $F^+$ and $F^-$ are the left and right
handed components of $R$.

The signs $s^+ = s^- = -1$ in (\ref{tetform+}) and (\ref{tetform-}) lead to the action $-I_{GR}$, 
which 
(in the absence of matter) is classically equivalent to $I_{GR}$. Opposite signs, $s^+ = -s^-$,
produce the actions $\pm I_1$. Finally, $\Sg^\pm$ having $V^+ + V^- = 0$, given by (\ref{degform}),
yield the actions $I_2$ and $I_3$. The matrix $T \in SO(3)$ in (\ref{degform}) is absorbed in an 
adjustment of the gauge of the right handed connection $A^-$. Put another way, the separate $SU(2)$ gauge
freedoms of $A^+$ and $A^-$ are reduced to a single gauge $SU(2)$ by fixing $T = \One$.
This leaves two possibilities: $s = 1$, corresponding to $I_3$, and $s = -1$, corresponding to
$I_2$.

\section{Solutions}  \label{solns}

The solutions of the theory defined by $I_\psi$ are generally solutions to one of the four
branch actions on all of spacetime. However, in special cases they consist of solutions 
belonging to different branches spliced together at transition hypersurfaces. In the following
the forms of the solutions to the branch actions $I_1$, $I_2$, and $I_3$ are outlined, and the
conditions under which $I_1$, $I_2$, and $I_3$ branch solutions can be spliced into a GR
solution are studied. The canonical formulation of the theory defined by $I_\psi$ is also
outlined, and it is shown that, generically, the left and right handed canonical data of a GR
solution does not have an alternative development into a solution of one of the other branches,
so the left and right handed data together specify the GR branch as well as the solution within
the GR branch. On the other hand, left handed GR data alone (such as Ashtekar's canonical variables),
while sufficient to specify a solution within the GR branch,  
always has alternative developments in the $I_1$ and $I_3$ branches.

The theories defined by $I_1$, $I_2$, and $I_3$ are extremely simple. 
The stationary points of $I_3$ satisfy the field equations
\bearr
F^{+\,i} & = & F^{-\,i}	\label{F+=F-}	\\
0 & = & D^\pm\wedge\Sg. \label{curlfreeI3}
\eearr
The most obvious class of solutions thus has $A^+ = A^-$ with $A^+$ and $\Sg$ chosen 
to be compatible in the sense that $D^+\wedge \Sg = 0$. Bengtsson \cite{Bengtsson} 
has solved this compatibility condition and obtained an expression for the connection 
in terms of $\Sg$, ${\Sg^*}_i^{\ag\bg} = \frac{1}{2} \eg^{\ag\bg\cg\dg}\Sg_{i\,\cg\dg}$,
the inverse of the $3\times 3$ matrix
\be
m_{ij} = \frac{1}{2}\eg^{\ag\bg\cg\dg}\Sg_{i\,\ag\bg}\Sg_{j\,\cg\dg},
\ee
and 
\be
t_{\ag\bg\cg\,ij} = 3\eg_{ijk}\di_{[\ag}\Sg_{\bg\cg]}^k. 
\ee
His solution is
\be			\label{Bengtsson_soln}
A^{+\,i}_\m = -2 t_{\m\bg\cg}{}^i{}_j{\Sg^*}^{\bg\cg}_k m^{-1\,jk} -
2 \Sg^i_{\m\bg}{\Sg^*}^{\bg\cg}_j t_{\cg\dg\sg\,km} {\Sg^*}^{\dg\sg}_n m^{-1\,mn}m^{-1\,jk}.
\ee
Of course this holds only provided that $m$ is invertible, which it is for generic $\Sg$.
Thus for generic $\Sg$ there
is always a corresponding solution and, since the compatibility of the connection with $\Sg$
defines the connection completely, (\ref{curlfreeI3}) implies that $A^+ = A^-$. There are
also solutions for which $A^+ \neq A^-$, but in these both $\Sg$ and $F^+ = F^-$ must
be algebraically special: If the matrix $n_{ij} = \eg^{\ag\bg\cg\dg}H_{i\,\ag\bg}H_{j\,\cg\dg}$
corresponding to any linear combination $H = a\Sg + b F^+$ is invertible then 
(\ref{curlfreeI3}) and the Bianchi identities imply that $A^+$ and $A^-$ are given by the same
expression in terms of $H$ and are thus equal.

At stationary points of $I_2$ $F^+ = -F^-$. One immediately sees that the fields must be
algebraically special in all the solutions: $D^-\wedge F^- = 0$ implies $D^- \wedge F^+ = 0$, 
just as for $I_3$ branch solutions. Thus, if $F^+$, $\Sg$, or any linear
combination satisfy the invertibility condition then $A^+ = A^-$, implying that $F^+$ equals
$F^-$ instead of $-F^-$. That is, in this branch the curvature is zero, or the invertibility
condition does not hold for any linear combination of the curl free fields $F^+$ and $\Sg$.
The most obvious class of solutions in this branch has $A^+ = - A^- = n^i A$, with $n^i$
an $SU(2)$ adjoint vector field, and $A$ a 1-form field. As in the $I_3$ 
branch, there are some additional solutions beyond these most obvious ones.

Stationarity of $I_1$ implies only one field equation:
\be 	\label{w_e_compatibility}
D\wedge[e_I\wedge e_J] = 0
\ee
This equation ensures that the connection $\omega$ is compatible with $e$ in the sense that
$D\wedge e = 0$.\footnote{
\noindent{\em Proof}: (\ref{w_e_compatibility}) can be expanded as
\be
[D\wedge e_I] \wedge e_J - [D \wedge e_J] \wedge e_I = 0.
\ee
This implies $[D\wedge e_I] \wedge e_J \wedge e_I = 0\ \forall J$. Thus $D\wedge e_I = K\wedge e_I$
for some 1-form $K$. Substituting this into (\ref{w_e_compatibility}) we find 
$0 = K\wedge e_I\wedge e_J$, which implies $K = 0$, and thus 
\be
D\wedge e_I = 0.
\ee
\QED}
The other ostensible field equation, which follows from stationarity with respect to $e$, is
\be
0 = e_J\wedge R^{IJ} = D\wedge D\wedge e^I,
\ee
and is implied by (\ref{w_e_compatibility}). Any $e^I_\m$ and compatible connection 
$\omega^{IJ}_\m$ will thus satisfy the field equations.

While both GR and $I_1$ branch solutions can be described in terms of $e$ and $\omega$
these fields are defined differently in terms of the fundamental variables $\Sg^\pm$,
$A^\pm$ in the two branches. In terms of the fundamental variables
the GR branch can be distinguished from the $I_1$ branch using the fact that
\be
\Sg^{+\,1}_{\mu[\ag}\Sg^{+\,2}_{\bg\cg}\Sg^{+\,3}_{\dg]\n} = \frac{s^-}{s^+}
\Sg^{-\,1}_{\mu[\ag}\Sg^{-\,2}_{\bg\cg}\Sg^{-\,3}_{\dg]\n}
\ee
with $\frac{s^-}{s^+} = 1$ in the GR branch and $-1$ in the $I_1$ branch.
Both sides of the equation are non-zero; 
$\Sg^{+\,1}_{\mu[\ag}\Sg^{+\,2}_{\bg\cg}\Sg^{+\,3}_{\dg]\n} 
= \frac{1}{8}V^+_{\ag\bg\cg\dg} g_{\m\n}$, where $g_{\m\n} = e^I_\m e_{I\,\n}$ is the 
metric in either branch. The GR branch can of course also be distinguished 
from the $I_2$ and $I_3$ branches because, unlike in the 
GR branch, $V^+ + V^- = 0$ in these branches.

Under what circumstances can one splice together a GR solution and a solution of another
branch? It turns out to be possible only in spacial cases. Evolution from the GR branch 
to the $I_1$ or $I_2$ branches cannot occur across a hypersurface
on which the vierbein is well defined and non-degenerate 
($det[e^I_\m] \neq 0$). Evolution from the GR branch to the $I_3$ 
branch can occur when the vierbein is non-degenerate, but requires a
condition on the curvature on the hypersurface of transition, which doesn't 
hold on any hypersurface in a generic solution of GR. 

To demonstrate these claims we need the junction conditions 
that spliced solutions must satisfy at the transition hypersurface. 
To derive these we first need to delimit the types of fields we are 
willing to consider as solutions to the field equations. Solutions 
will be required to be piecewise differentiable, with the set of points 
of non-differentiability a union of differentiable submanifolds. 
Moreover, at 3-surfaces of non-differentiability in the interior of the 
spacetime the fields are required to have well defined, finite limiting 
values when the 3-surface is approached from either side, and parallel 
transport along any differentiable curve is required to depend continously 
on the curve. Requiering the points of non-differentiability to form 
differentiable manifolds makes it possible to replace the differential 
field equations by junction conditions at these manifolds. The requirement 
that the fields have well defined limits at 3-surfaces in the interior does
not disallow the development of singularities in the fields (which would
certainly exclude physically interesting solutions) but simply states
that we will not develop solutions through such singularities. Of course 
singularities might develop on hypersurfaces that are do not cut the 
spacetime in two.\footnote{Such surfaces can still be considered as excluded 
from the spacetime so that they do not form part of its interior.}
The stationarity of the action will require some sort of junction 
condition to hold across these surfaces but we will not need these. Finally, 
allowing the parallel propagator along a curve 
to depend discontinously on the curve would lead to ambigous field equations in 
some cases.

Now suppose we slice an open region of spacetime with trivial topology into 
differentiable hypersurfaces $H_t$, continously parametrized by a ``time" 
$t \in \Real$. Let's also adopt adapted coordinates $x^\m$ with $x^0 = t$, 
and let latin indices $a,b,c,... \in \{1,2,3\}$
so that they indicate ``spatial" components in the tangent hyperplanes to 
$H_t$. It follows immediately from the restrictions placed on the connection
that $A^{\pm\,i}_a$ is continous in time, $t$, almost everywhere: The 
parallel propagator along a spatial curve, kept fixed in the coordinates
$x^a$, is a continous function of time. Moreover, our hypothesies imply 
that almost everywhere on $H_t$ $A^\pm$ has well defined limiting values as 
$H_t$ is approached from either side. Thus the integrals of the two limits
of the connection on $H_t$ along any differentiable curve are equal, implying 
that the limits themselves are equal almost everywhere.

The other quantity that must be continous in $t$ (almost everywhere on
$H_t$) is $\Sg^\pm_{i\,ab}$. This can be seen by considering the integrals 
of the field equations $D^\pm\wedge\Sg^\pm = 0$ over infinitesimal three
dimensional ``Gaussian boxes". For a connection of the form we are allowing 
it is possible to choose a gauge so that the connection vanishes at a point 
inside a given box. In this gauge the connection terms appearing in the
covariant curls $D^\pm\wedge\Sg^\pm$ can be neglected inside the box as it
is shrunk to infinitesimal size, and Stoke's theorem can be applied. 
Choosing suitable boxes it is then easy to show that the integrals of
$\Sg^\pm$ over any spatial 2-surface which is fixed in the coordinates
$x^a$ is continous in $t$, and the existence of the the two one sided
limits of $\Sg^\pm$ on $H_t$ then implies
that $\Sg^\pm_{i\,ab}$ is continous in $t$ almost everywhere on $H_t$.

The continuity of $A^{\pm\,i}_a$ and $\Sg^\pm_{i\,bc}$, or equivalently 
$\tilE^{\pm\,a}_i = \eg^{abc} \Sg^\pm_{i\,bc}$, can also be
understood in terms of a canonical formulation of the theory. 
The action (\ref{LR_action}) can be rewritten with the spatial and time 
components explicitly separated:
\bearr	
I_\psi & = & \int dt \int d^3x\ \dot{A}^{+\,i}_a \tilE^{+\,a}_i 
- \dot{A}^{-\,i}_a \tilE^{-\,a}_i
+ A^{+\,i}_0 D^+_a \tilE^{+\,a}_i - A^{-\,i}_0 D^-_a \tilE^{-\,a}_i 
\nonumber\\
&&
\ \ +\: 2\Sg^+_{i\,0a} \tilde{B}^{+\,ia} - 2\Sg^-_{i\,0a} \tilde{B}^{-\,ia}
+ \tilde{\theta}^{ab}[\Sg^{+\,i}_{0a}\Sg^+_{i\,0b} - \Sg^{-\,i}_{0a}\Sg^-_{i\,0b}]
\nonumber \\
&&
\ \ +\: \chi^a{}_b [\Sg^{+\,i}_{0a}\tilE^{+\,b}_i - \Sg^{-\,i}_{0a}\tilE^{-\,b}_i] 
+ \subtil{\tau}_{ab}  [\tilE^{+\,ia}\tilE^{+\,b}_i - \tilE^{-\,ia}\tilE^{-\,b}_i].	\label{canon_action}
\eearr
The lagrange multipliers $\tilde{\theta}$ and $\subtil{\tau}$ are symmetric 
while $\chi$ is 
trace free. ($\tilde{\theta}^{ab} = 4\psi^{0a0b}$, 
$\chi^a{}_b = 2\psi^{0acd}\eg_{bcd}$, and $\subtil{\tau}_{ab} = \frac{1}{4}
\psi^{cdef}\eg_{acd}\eg_{bef}$). $\tilde{B}^{\pm\,ia} = \eg^{abc}F^{\pm\,i}_{bc}$ 
is the left/right $SU(2)$ magnetic field. One sees at once that the canonical 
coordinates are the $A^{\pm\,i}_a$ and their conjugate momenta, 
$\pm\tilE^{\pm\,a}_i$. 
The field equations consist of constraints and expressions for the $t$ 
derivatives of $A^{\pm\,i}_a$ and $\tilE^{\pm\,a}_i$ as functions of both the 
spatial and time components of $\Sg^\pm$, $A^\pm$, and the curvatures $F^\pm$. 
appearing in the action. By our hypothesies on the fields these expressions 
must be finite, so the $t$ evolution
of $A^{\pm\,i}_a$ and $\tilE^\pm$ is continous. It is clear from this point
of view that there are no further, independent, requierments of continuity
in $t$, since there are no further equations connecting the fields at
different $t$.

Now let's see how the continuity conditions restrict transitions from the GR 
branch to other branches. Let's first consider transitions to the $I_1$ 
branch. Let $B$ be the hypersurface separating the GR and $I_1$ phases and 
choose a piecewise differentiable slicing $H_t$ so that $H_0$ coincides with 
$B$ wherever $B$ is differentiable. In particular this means that $H_0$ 
coincides with $B$ whenever the fields are non-differentiable at 
$B$.\footnote{
It would of course be simpler if one could identify all of $B$ with $H_0$,
but it is not clear to the author that all of $B$ is even piecewise differentiable.}
If the $SO(4) = SU(2)_L\otimes SU(2)_R$ gauge is chosen so that $e^0_a = 0$ 
the expressions (\ref{tetform+}) and (\ref{tetform-}) for $\Sg^\pm$ imply
\be		\label{dtriad}
\tilE^{\pm\,a}_i = s^\pm e^a_i det[e^j_b]
\ee
where $e^a_i$ is the inverse of the $3\times 3$ matrix $e^j_b$. 
From (\ref{dtriad}) follows the gauge invariant result that
\be
det[\tilE^{\pm\,a}_i] = s^\pm det[e^j_b]^2.
\ee
Thus $det\tilE^+ = \frac{s^+}{s^-} det\tilE^-$.

In the GR branch $s^+ = s^-$ while in the $I_1$ branch $s^+ = -s^-$, so to 
cross from one branch to another requires that $det\tilE^+$ or 
$det\tilE^-$ changes sign. The continuity of these determinants thus 
implies that one of them (and thus also the other) is zero on 
$B$. This in turn implies that $det[e^i_a] \rightarrow 0$ as $B$ is approached,
so $e^I{}_\m$ does {\em not} approach a non-degenerate limiting vierbein
on $B$.

In the $I_2$ branch (in a suitable gauge) 
$\Sg^+ = -\Sg^-$ so $det\tilE^+ = - det\tilE^-$, as in the $I_1$ branch. 
Therefore, just as in the $GR \rightarrow I_1$ transition, 
$det[e^I_\m] = 0$ on $B$.

What about a transition to the $I_3$ branch? In this branch 
$\tilE^{+\,a}_i = \tilE^{-\,a}_i$ in an appropriate gauge, just as in the GR
branch in the gauge in which $e^0{}_a = 0$ (recall (\ref{metric_form})). Therefore it 
seems to be possible to make the transition with $e^I{}_\m$ approaching
a well defined and non-degenerate limiting vierbein as $B$ is approached
from the GR phase. We will see that this is in fact the case provided the
curvature satisfies a constraint. Let's therefore suppose that $det[e^I{}_\m]
\neq 0$ on an open subset $U$ of $B$. In the $I_3$ phase $V = \frac{1}{2}
(V^+ + V^-) = 0$, while in the GR phase it is $V = det[e^I{}_\m]\eg \neq 0$,
so $\Sg^+$ or $\Sg^-$ must be discontinous on $U$. It follows that $U$ is
piecewise differentiable and coincides with $H_0$. 

In the $I_3$ phase $F^+ = F^-$ in the gauges in which $\Sg^+ = \Sg^-$.
Thus the continuity of $A^{\pm\,i}_a$ implies that $F^{+\,i}_{ab}
= F^{-\,i}_{ab}$ on $U$ in the gauge $e^0{}_a = 0$. This
is actually a sufficient condition for the existence of an $I_3$ branch
solution matching the GR solution on $B$: provided this condition holds
$A^\pm$ and $\Sg^\pm$ constant in $x^0$, with $A^{\pm\,i}_0 = 0$ and 
$\Sg^\pm_{i\,0a} = 0$, solve the $I_3$ field equations.

Reconstructing the $SO(4)$ curvature from $F^\pm$ we find that the constraint 
$F^+ = F^-$ on the curvature is equivalent to $R^{0i}{}_{ab} = 0$ on $U$. 
It is thus 
possible to evolve from the GR phase to the $I_3$ phase without the vierbein 
becoming degenerate at the transition iff there exists a hypersurface in the 
GR solution on which $R^{0i}{}_{ab} = 0$. No such hypersurface exists in a 
generic solution to GR. The vacuum Einstein equations reduce the Riemann curvature
to the Weyl curvature. $R^{0i}{}_{ab} = 0$ places five constraints on the Weyl
curvature on the transition hypersurface. Two of these can be taken to be the
vanishing of the two algebraically independent scalar densities\footnote{
The algebraic independence of $a$ and $b$ can be seen much more clearly when 
they are expressed in terms of the left and right handed Weyl curvature 
spinors. In a
solution the left and right handed curvatures can be written as
\be		\label{phiform}
F^{\pm\,i}_{\m\n} = \phi^{\pm\,ij}\Sg^\pm_{j\,\m\n},
\ee 
where $\phi^+$ and $\phi^+$ are traceless, symmetric $3\times 3$ matrices.
({\em Outline proof}: $D^\pm \wedge \Sg^\pm = 0 \implies D \wedge e = 0$ $\implies
R^{\ag}{}_{[\m\n\sg]} = 0 \implies \psi^{\sg\ag}{}_\sg{}^\bg = 0$ $\implies \Sg^{+\,i}_{\ag\bg}
\psi^{\ag\bg\cg\dg}\Sg^{-\,j}_{\cg\dg} = 0$. This implies (\ref{phiform}), and the tracelessness
of $\phi^\pm$ follows straightforwardly from $\psi^{[\ag\bg\cg\dg]}= 0$).
$\phi^\pm$ is the left (right) handed Weyl curvature spinor expressed as an $SO(3)$ 
tensor. Clearly our condition that the spatial components of the left and right handed
curvatures be equal in the gauge in which $\tilE^+ = \tilE^-$ is equivalent to $\phi^+
= \phi^-$. In terms of the $\phi$ $a = (tr[\phi^{+\,2}] - tr[\phi^{-\,2}])\sqrt{g}$ and
$b = (\det\phi^+ - \det\phi^-)\sqrt{g}$, modulo constant factors. From these forms it is 
easy to check algebraic independence.}
\be
a = R^{IJ}{}_{\ag\bg}R_{IJ\,\cg\dg} \eg^{\ag\bg\cg\dg}
\ee
and
\be
b = R_I{}^J{}_{\m\ag} R_J{}^K{}_{\bg\cg} R_K{}^I{}_{\dg\n} g^{\m\n} \eg^{\ag\bg\cg\dg}.
\ee
Using linear perturbation theory one can verify that any value of the Weyl 
curvature at a single point can be realized in a solution. That is to say, 
the consistency of the Einstein equations does not require any algebraic 
restriction on the Weyl curvature. Thus if $a$ or $b$ are zero at a point 
they will be perturbed to a non-zero value in a
neighborhood of that point by a generic piecewise differentiable perturbation. 
A generic superposition of weak perturbations (i.e a generic small perturbation) will
produce $a$ and $b$ that vanish only on some hypersurfaces. Now, the condition that
$R^{0i}{}_{ab} = 0$ on the transition hypersurface requires that {\em both} $a$ and $b$
vanish on that hypersurface. But it is easy to see that a generic perturbation will separate
the $a = 0$ and $b = 0$ hypersurfaces: Any perturbation can be combined with a 
diffeomorphism so that, at least in an open neighborhood, the $a = 0$ surface is preserved.
The perturbations obtained in this way are in fact all the perturbations that preserve
the $a = 0$ surface. Thus, since $a$ and $b$ are algebraically independent, these perturbations
will generically make $b \neq 0$ at some point of the $a = 0$ surface, in fact they will make
$b \neq 0$ everywhere on this surface except on some 2-surfaces. It follows that
in a generic solution of GR there is no hypersurface on which $R^{0i}{}_{ab} = 0$ everywhere.

In terms of the canonical framework the preceeding results imply that complete 
left handed {\em and} right handed canonical data $(A^\pm, \tilE^\pm)$ corresponding
to a non-degenerate solution of GR is not initial data for any $I_1$ or $I_2$ branch
solution, and forms the initial data of an $I_3$ branch solution only in the exceptional
case in which $F^{+\,i}_{ab} = F^{-\,i}_{ab}$ in the gauge in which $\tilE^{+\,ia}
= \tilE^{-\,ia}$.\footnote{
In the exceptional case that the data is consistent with both
a GR solution and an $I_3$ solution the branch must be selected by a suitable choice of
the Lagrange multipliers $\Sg^\pm_{i\,0a}$.}

In GR the left handed part of the canonical data is sufficient to specify the 
whole solution. The conditions that define the GR branch (namely that the 
4-volume density $2[\Sg^{+\,i}_{0a}\tilE^{+\,a}_i - \Sg^{-\,i}_{0a}\tilE^{-\,a}_i]$ 
not vanish and that $det\tilE^+$ and $det\tilE^-$ are non-zero
with the same sign) and the constraints on the canonical variables and lagrange
multipliers following from (\ref{canon_action}) imply that in a suitable gauge
\bearr
\tilE^{-\,a}_i & = & \tilE^{+\,a}_i	\\
A^{-\,i}_a & = & - A^{+\,i}_a + \eg^i{}_{jk}\omega^{jk}(\tilE^+)
\eearr
where $\omega^{jk}_a$ is the spatial spin connection compatible with the dreibein
$e^a_i = \tilE^{+\,a}_i/\sqrt{det\tilE^+}$. The constraints also determine the
Lagrange multipliers $A^-_0$, $\Sg^-_{i\,0a}$, $\theta$, $\chi$, and $\tau$ in terms
of $A^+_0$ and $\Sg^+_{i\,0a}$ in the same gauge, and thus allow the computation
of the evolution of the canonical variables from the left handed Lagrange
multipliers.

However, other right handed data can be chosen to match the given left handed data so
that the development is an $I_1$ or $I_3$ branch solution. We may choose 
\bearr
\tilE^- & = & - \tilE^+	\\
A^{-\,i}_a & = & - A^{+\,i}_a + \eg^i{}_{jk}\omega_a^{jk}(\tilE^+)
\eearr
to obtain $I_1$ initial data, or
\bearr
\tilE^- & = & \tilE^+	\\
A^- & = & A^+ 
\eearr
to get $I_3$ initial data. (Only in special cases is it possible to complete left handed
data to form $I_2$ data, because $F^+ = - F^-$ requires that $-F^+$ is a curvature of some
connection which is true only for certain special connections $A^+$).

\section{A comment on quantizing the model}	\label{comment}

A fairly straightforward path integral 
quantization of the classical theory (\ref{LR_action}) defined by the principle 
``left-handed area = right-handed area" on a simplicial spacetime leads to
the Barrett-Crane model.\footnote{
The author hopes to present the details in a forthcoming publication.}
Our study of the classical solutions however suggests a potential problem with
such a path integral. It seems that semiclassically the path integral is 
completely 
dominated by histories near $I_3$ branch solutions, so that the quantum theory
is a quantization of the $I_3$ branch only, and not the GR branch. The argument
that leads to this conclusion runs as follows:
Consider a formal path integral over the continuum fields,
and take $\tilE^\pm$ as the boundary data to be kept fixed in the path 
integral.
In a semiclassical situation the path integral should be dominated by histories
near classical solutions. Suppose that the boundary data is compatible with a 
classical GR solution and, for notational convenience choose the gauge in 
which $\tilE^+ = \tilE^-$. In metric language we have fixed the 3-metric of 
the boundary
and for the vierbein we have adopted the gauge in which $e_0^\m$ is normal to 
the boundary). This boundary data will generically admit only one 
4-diffeomorphism
equivalence class of solutions to GR. But now consider $I_3$ solutions. Choose
any $\Sg^+$ such that $\Sg^{+\,i}_{ab} = \frac{1}{2}\eg_{abc}\tilE^{+\,ic}$ 
on the boundary. Whatever the boundary values are, almost all the spacetime 
fields $\Sg^+$ will define an invertible matrix 
$m_{ij} = \frac{1}{2}\eg^{\ag\bg\cg\dg} \Sg^+_{i\,\ag\bg}\Sg^+_{j\,\cg\dg}$, 
and will thus define via (\ref{Bengtsson_soln}) a unique connection $A^+$ 
satisfying the compatibility condition $D^+\wedge \Sg^+ = 0$. 
By choosing $\Sg^- = \Sg^+$ and $A^- = A^+$ one obtains a complete $I_3$ 
solution matching the boundary data. Thus there is one solution for almost 
any $\Sg^+$ satisfying the boundary data. This set of solutions is much 
larger than the 4-diffeo equivalence class of one solution. In GR too all the 
fields in a solution are determined by $\Sg^+$, up to an $SU(2)_R$ gauge freedom.
For instance, the metric is given by Urbantke's formula \cite{Urbantke,Bengtsson}
\be			\label{Urbform}
\sqrt{g}g_{\m\n} = 8\eg^{\ag\bg\cg\dg}\Sg^{+\,1}_{\m\ag}\Sg^{+\,2}_{\bg\cg}
\Sg^{+\,3}_{\dg\n}.
\ee
But in GR only a very restricted set of $\Sg^+$ histories correspond to solutions.
If the 4-diffeo and $SU(2)_L\otimes SU(2)_R$ gauge freedom is fixed, for instance
by the harmonic condition $\di_\m g^{\m\n} = 0$ on the inverse of the left-handed
Urbantke metric (\ref{Urbform}), and $A_0^{+\,i} = 0$, 
$\Sg^{+\,i}_{ab} = \Sg^{-\,i}_{ab}$, then there will be only one GR solution matching 
the boundary data but an infinity of $I_3$ solutions. This suggests that in semiclassical
situations the path integral is completely dominated by the $I_3$ branch 
solutions, so that the quantum theory defined by the path integral is {\em not} a 
quantization of GR. 

\section{Conclusion}	\label{conclusion}

The constraint ``left-handed area = right-handed area" leads to a nice 
geometrical, and intuitively accesible, prescription for obtaining general 
relativity from BF theory. However, the resulting theory has several phases, 
only one of which reproduces GR. Moreover, some of the other phases contain 
in a sense many more solutions. GR is in this theory a sort of degenerate 
phase, analogous to unstable equilibria in mechanics, which classically can 
persist forever provided the initial data are taken from a special set
of measure zero. Thus
one has to take care when building a quantum theory of GR from BF theory and 
``left-handed area = right-handed area" in order that the resulting theory is
not simply the quantum theory of one of the other phases.

\section{Acknowledgements}

I am obliged to Rodolfo Gambini, Carlo Rovelli, and Jose Zapata for helpful 
discussions and encouragement. This work was supported by the Comision
Sectorial de Investigaci\'on Cientifica of the Universidad de la Rep\'ublica.

\end{document}